\documentclass[aps,pre,twocolumn,byrevtex,superscriptaddress]{revtex4-1}

 \usepackage{epsfig}
 \usepackage{color}
 \usepackage{amsmath}
 \usepackage{amsthm}
 \usepackage{amsfonts}
 \usepackage{amssymb}

 \usepackage[english]{babel}
 \usepackage[utf8]{inputenc}

\newtheorem{theorem}{Theorem} 
\DeclareMathOperator{\sech}{sech}

\DeclareMathOperator{\arctanh}{arctanh}

\def\be{\begin{equation}}
\def\eea{\end{eqnarray}}
\def\ee{\end{equation}}
\def\bea{\begin{eqnarray}}
\def\ea{\end{array}}
\def\ba{\begin{array}}

\newcommand{\bel}[1]{\begin{equation}\label{#1}}

\newcommand{\bs}{\backslash}

\def\zzz{{\mathchoice {\hbox{$\sf\textstyle Z\kern-0.4em Z$}}
{\hbox{$\sf\scriptstyle Z\kern-0.3em Z$}}
{\hbox{$\sf\scriptscriptstyle Z\kern-0.2em Z$}}
{\hbox{$\sf\textstyle Z\kern-0.4em Z$}}}}

\begin{document}

\title{
  Correspondence between spanning trees and the Ising
  model on a square lattice}

\author{G. M. Viswanathan}

\affiliation{Department of Physics
and \mbox{National Institute of Science and Technology of Complex Systems,}
Universidade Federal do Rio Grande do Norte, 59078-970
Natal--RN, Brazil}

\begin{abstract}
{    An important problem in statistical physics concerns the
  fascinating connections between partition functions of lattice
  models studied in equilibrium statistical mechanics on the one hand and
  graph theoretical enumeration problems on the other hand.
We investigate the nature of the relationship between the number of
spanning trees and the partition function of the Ising model on the
square lattice.}
  The spanning tree generating function $T(z)$ gives the spanning tree
  constant when evaluated at $z=1$, while giving the lattice green
  function when differentiated.  It is known that for the infinite
  square lattice the partition function $Z(K)$ of the Ising model
  evaluated at the critical temperature $K=K_c$ is related to $T(1)$.
  Here we show that this idea in fact generalizes to all real temperatures.
    We prove that
\mbox{$ ( Z(K)
\sech 2K ~\!)^2 = k
\exp\big[
T(k)
\big]
$}\,,
where $k= 2 \tanh(2K) \sech(2K)$.
The identical Mahler measure connects the two seemingly disparate
quantities $T(z)$ and $Z(K)$.  In turn, the Mahler measure is
determined by the random walk structure function.  Finally, we show
that the the above correspondence does not generalize
in a straightforward manner 
to non-planar
lattices.
  
\end{abstract}

\maketitle

\section{Introduction}

There is continued interest in exploring the fascinating connections
between partition functions of lattice models in statistical physics
on the one hand and graph theoretical enumeration problems on the
other.  In the 1960s, the seminal chapter by Kasteleyn in Harary's
{\it Graph Theory and Statistical Physics} \cite{gttp} led to
widespread recognition of the effectiveness of graph enumeration
methods for solving combinatorial problems in statistical mechanics.
In the 1970s, Wu showed that the number of spanning trees is related
to the partition function of ice-type models~\cite{wu77}. On the
square lattice, for example, it is known that the spanning tree
constant is related to the partition function of the critical Ising
model, since both can be expressed in terms of Catalan's constant.  By
the late 1980s, however, interest in exactly solvable models had to
some extent become overshadowed by the appeal of concepts such as
universality, scaling, and the renormalization
group~\cite{stanley-rmp1999,fisher-rmp1998}, which led to advances not
only within the field of statistical physics but more generally in the
interdisciplinary study of complex systems~\cite{csbook1,csbook2} and
complex networks~\cite{cnbook1}.
Nevertheless, there remain questions
concering exactly solvable models in statistical mechanics (and their
graph theoretical analogs) that cannot be tackled using scaling laws
and related methods.

Here we revisit the spanning tree problem and
ask whether or not the relationship between spanning tree enumeration
and the Ising model can be generalized to non-critical temperatures.
This question is (almost certainly) impossible to answer using
renormalization group methods or scaling analysis, due to breakdown of
scale invariance symmetry far from critical points.  We instead take a
different approach. In a sense that we make precise below, we directly
compare the generating functions related to certain types of graphs,
viz., spanning trees and multipolygons.

{   
  The main results we report here represent an advance for the
  following reasons: (i) Theorem 1 below sheds light on how the
  partition function of the Ising model on the square lattice is
  related to a corresponding generating function for spanning trees,
  via the random walk structure function and associated Mahler
  measures; (ii) we report preliminary results that strongly suggest
  that the theorem generalizes to other planar lattices; and 
  finally  (iii) our results have a bearing  on the question  of 
  why the Ising problem
  is notoriously difficult for non-planar lattices. 
    }

In Section II we report the key advances. In Section III we
interpret the results in terms of 
the underlying  physics. Finally in Section IV we  discuss
possible generalizations.

\section{Method and Results}

We first review the definition of the spanning tree constant.  Trees
are connected graphs that contain no loops (i.e., polygons).  A
spanning tree on a given graph is a tree that connects every node on
that graph.  The number $n_{\mathcal L}$ of spanning trees on a
regular lattice $\mathcal L$ grows exponentially in the number $N$ of
lattice sites. Hence, the following limit is well defined:

\be \lambda_{\mathcal L}= \lim_{N\to \infty} \frac {\log n_{\mathcal
    L}(N)} N ~.  \ee
The number $\lambda_{\mathcal L}$ is known as the spanning tree
constant for the lattice ${\mathcal L}$.

The spanning tree constant also bears a connection with lattice
models.  Aiming to study phase transitions, Fortuin and Kasteleyn in
the 1970s introduced a model that covers many others as special cases,
including the Ising model, the Ashkin-Teller-Potts model and
percolation \cite{fortuin-kasteleyn}.  Using this approach, Wu
subsequently showed that $\lambda_{\mathcal L}$ is expressible in
terms of a partition function of a lattice model, and thereby was able
to calculate the exact spanning tree constant on several planar
lattices \cite{wu77}.
In what follows, we adopt some of the notation of
ref.~\cite{arXiv:1207.2815}.
Let $\Lambda_{\mathcal L}$ denote the random walk structure function
for the lattice $\mathcal L$, i.e., the Fourier transform of the
discrete step probability distribution, and let ${\rm z}_{\mathcal L}$
denote the coordination number of the lattice. Then the spanning tree
constant of a $d$-dimensional regular lattice can be expressed as

\be
\lambda= \log 
{\rm z}_{\mathcal L}
+ 
\frac 1{ (2\pi)^d}
\int_{-\pi}
^\pi
\ldots
\int_{-\pi}
^\pi
\log
    [1-\Lambda_{\mathcal L}(\vec k)]
    ~dk_1\ldots dk_d
~.
    \ee
Specializing to the infinite square lattice, the random walk structure
function is given by
\bel{eq-def-Lambda}
\Lambda(\vec k) = \tfrac 1 2 (\cos k_1 + \cos k_2)
~.
\ee
  Hence,   on the square lattice, the spanning tree
constant is given by
\begin{align}
\lambda&= \frac 1{ 4\pi^2}
\int_{-\pi}
^\pi
\int_{-\pi}
^\pi
\log
[4 - 2(\cos k_1 + \cos k_2)] ~dk_1dk_2
\\
&={4G\over \pi} \approx 1.166 \label{eq-spc-G}
\end{align}
where $G\approx 0.915$ is Catalan's constant.

In 1944, Onsager had already shown that Catalan's constant appears in
the expression for the partition function of the critical Ising
model~\cite{ons}.  Let $Z(K)$ denote the partition function of the
Ising model on the infinite square lattice:
\begin{align}
\log Z(K)
&= \log (2\cosh 2K ) \nonumber
\\
& 
\quad
+
\frac 1 {8 \pi^2}
\underset{-\pi~}{\overset{~\pi} {\iint}}
\log [1-\tfrac k 2 (\cos k_1 + \cos k_2)]
~dk_1 dk_2 \label{eq-def-Z}
\end{align}
where
\bel{eq-def-k}
k= 2 \tanh 2K  \, \sech 2K
~.
\ee
The critical temperature $K_c=\arctanh(\sqrt 2 -1)$ was first found by
Kramers and Wannier~\cite{kw}, and corresponds to the condition $k=1$.
Subtituting $K=K_c$ into (\ref{eq-def-Z}) gives us
\bel{eq-Z-G}
\log Z(K_c)=\frac 1 2 \log 2 + \frac {2G} {\pi}
~.
\ee

Eliminating Catalan's constant from (\ref{eq-spc-G}) and
(\ref{eq-Z-G}), we find that $\lambda$ and $Z(K_c)$ are related by
\bel{eq-crit-eq}
[Z(K_c)]^2 = 2 e^{\lambda}
~.
\ee
This result is of course not new. Indeed, it is well known to
specialists that the critical Ising model is related to the problem of
enumerating of spanning trees (see, e.g., ref.~\cite{deTiliere01} and
Eq.~(57) in ref.~\cite {arXiv:1207.2815}).

The question we address here is whether the above relation holds only
at the critical point or whether it generalizes in some way.
Specifically, we ask here whether or not (\ref{eq-crit-eq}) can be
extended to non-critical temperatures.  It is well known that the
partition function of the Ising model is closely related to the
generating function of ``loops'' or multipolygons on the square
lattice, i.e. those graphs all whose nodes have even degree and whose
edges connect only nearest neighbors.  The critical temperature $K_c$
correponds to weighting the edges of multipolygons in such a way that
multipolygons of small and large size make comparable overall
contributions.
The form of 
Eq. (\ref{eq-crit-eq}) strongly suggests 
that spanning tree
constant is related to some ``spanning tree generating function''
evaluated at a special critical value of some tunable parameter.  This
generating function must generalize the spanning tree constant, such
that when evaluated at other values of the parameter, it should be
related to (\ref{eq-def-Z}) with \mbox{$K\neq K_c$.}  What is the
correct or ``natural'' way to define this spanning tree generating
function that generalizes the spanning tree constant?  Fortunately,
the answer is known \cite{rosengren,arXiv:1207.2815}.

The spanning tree generating function $T_{\mathcal L}(z)$ was defined
by Guttmann and Rogers \cite {arXiv:1207.2815} as

\be
T_{\mathcal L}(z)
=
\log {\rm z}_{\mathcal L}+
\frac{1}{(2\pi)^d }
\!
\int_{-\pi}^\pi \!\!\!\!
dk_1
\ldots\!
\int_{-\pi}^\pi \!\!\!
\!
dk_d
~\log [1/z - \Lambda_{\mathcal L}(\vec k)]
~.
\ee
Clearly,  $T_{\mathcal L}(1)=\lambda_{\mathcal L}$. 
Moreover, differentiating  one obtains

\be
-z \frac {d T_{\mathcal L}(z) }{dz}
=
\frac{1}{(2\pi)^d }
\int_{-\pi}^\pi dk_1
\ldots
\int_{-\pi}^\pi dk_d
~\frac 1 {1 - z \Lambda_{\mathcal L}(\vec k)}
~.
  \ee
This integral is the Lattice Green function~\cite{lgf}, whose value at
$z=1$ is related to the probability of a random walker to return to
the origin.

On the square lattice, 
the spanning tree generating function is given by 

\bel{eq-def-T}
T(z)=
\log 4 \! +\!
\frac{1}{4\pi^2  }\!
\underset{-\pi~}{\overset{~\pi} {\iint}}
\!\log [\tfrac 1 z  - \tfrac 1 2 (\cos k_1 + \cos k_2) ]
dk_1 dk_2
~.
  \ee

  The integral that appears in the above expression is, up to 
  constants, identical 
  to the
one in (\ref{eq-def-Z}).
It thus follows that $
  \log ( Z(K) \sech 2K)
  = \tfrac 1 2 (T(k) + \log k ) 
~,
  $
  from which we immediately obtain the following new result:

  \begin{theorem}  Let $T(z)$ be the
the spanning tree generating
  function given by (\ref{eq-def-T}) for the infinite square lattice and let
$Z(K)$    
be
  the partition function
  of the Ising model $Z(K)$ given by (\ref{eq-def-Z}). Then, for all $K \in \Bbb R$,  
the following identity holds,
\bel{eq-theorem1}
\big(Z(K)
\sech 2K\big)^2 = k
\exp\big[
T(  k) 
\big]
~, 
\ee
with $k$ given by (\ref{eq-def-k}).
\label{th-main}
\end{theorem}

The identity  (\ref{eq-theorem1}) is the correct generalization of
(\ref{eq-crit-eq}) to non-critical temperatures $K\neq K_c$.  
We originally arrived at Theorem \ref{th-main}
by noting that $T(z)$ and $Z(K)$
can both be written in terms of the same $_4 F_3 $ generalized
hypergeometric function. See the appendix for details.

\section{Discussion}

The physics that underpins Theorem \ref{th-main} can be understood in
terms of random walks. We briefly discuss this relationship.  The integrals
that appear in the expressions for $T(z)]$ and $Z(K)$ can be written
  in terms of a Mahler measure (for $T$ and  $Z$ real). The (logarithmic) Mahler measure of a
  Laurent polynomial $P(z_1,\ldots,z_n ) $ is conventionally defined
  according to
  
  \bel{eq-mahler-m-ererg}
\rm   m[P]= \int_0^1   \ldots \int_0^1  \log \big |
  P(e^{2 \pi i \theta_1},\ldots,
  e^{2 \pi i \theta_n})
  \big |
  ~d\theta_1\ldots d\theta_n
~.
  \ee
  If we choose  $P$ to be the symmetric 2-variable Laurent
  polynomial 
  \be
  p(\zeta)= \zeta +  z_1 + z_1^{-1} +  z_2 + z_2^{-1}
  \ee
  where $\zeta$ is not integrated over, we obtain the required integral:
  \be
     {\rm m}[p(\zeta)] =
     \frac 1 {4 \pi^2} 
     \underset{-\pi~}{\overset{~\pi} {\iint}}
     \log[\zeta + 2(\cos k_1 + \cos k_2) ] dk_1 dk_2
~~.
     \ee
  For
example, we find $T(z)={\rm m}[p(4/z)]$ for $z \in \Bbb R \bs \{0\}$.
The reason the identical Mahler measure is found in both $T(z)$ and
$Z(K)$ is that they can both be calculated in terms of a random walk
on  the lattice, as is well known. The polynomial
$p(\zeta)$ is in fact determined by the random walk structure function
$\Lambda(\vec k)$ for the square lattice:

\be
p(\zeta) = \zeta + 4 \, \Lambda(\vec z)
~,
\ee
with $\Lambda$ given by (\ref{eq-def-Lambda}) and $\vec z$ denoting
the vector with components $z_1$ and $z_2$. Hence, the connection
between the spannng tree generating function and the Ising model is
ultimately due to the fact that both share a mathematical relationship
with the random walk structure function for the square lattice.

\section{Conclusion}

We have generalized the relation between the spanning tree generating
function and the partition function of the Ising model to all real
temperatures.  We recover the known relation between the spanning tree
constant and the critical Ising model as a special case of our more
general result.  Does Theorem 1 generalize to other lattices?  For
planar lattices, one expects similar results to hold (and we hope to
systematically investigate such lattices when time permits).
{   
  As evidence in support of this idea,
we cite the the
triangular lattice.  The partition function $Z_{\mbox{\tiny Tri}}$ of the Ising model on the triangular 
lattice is known to be given by~\cite{montroll}, 
\begin{align}
\log ( \tfrac 1 2 
Z_{\mbox{\tiny Tri}} )
&=
\frac 1 {8 \pi^2}
\underset{0~}{\overset{~2\pi} {\iint}}
\log \bigg\{ 1+
  \cosh^3 2K  \nonumber
  \\&\quad\quad\quad
  - \sinh^2 2K (\cos k_1 + \cos k_2)
  \\&\quad\quad\quad \nonumber
  - \sinh^2 2K \cos (k_1 + k_2)
\bigg\} ~dk_1 dk_2
    ~.
\end{align}
The spanning tree generating function for this lattice is given by
\cite{arXiv:1207.2815},
\begin{align}
T_{\mbox{\tiny Tri}} &= \log 6
+  \nonumber
\frac 1 {8 \pi^2}
\underset{-\pi~}{\overset{~\pi} {\iint}}
\log \bigg\{ \frac 1 z - \frac 1 3
\big(\cos k_1 + \cos k_2
\\ &
\quad\quad\quad\quad\quad\quad\quad\quad\quad\quad
+ \cos (k_1+k_2)\big)
\bigg\}~ dk_1 dk_2
~.
\end{align}
Note that if we restrict $z$ and $K$ to real values, then we can
attempt to express both $Z_{\mbox{\tiny Tri}} $ and $T_{\mbox{\tiny Tri}} $
in terms of the same Mahler measure. This is possible by noting that
both both integrals can be written in terms of the identical structure
function 
\be
\Lambda_{\mbox{\tiny Tri}} (\vec k)
= \frac 1 3 (\cos k_1 + \cos k_2 + \cos (k_1+k_2) )
~.
\ee
A suitable  Mahler measure is obtained by choosing $P$ to be the
Laurent polynomial
\be
p_{\mbox{\tiny Tri}} (\zeta) = \zeta
+ z_1 + z_1^{-1} 
+ z_2 + z_2^{-1}
-z_1z_2 -(z_1z_2 )^{-1}
\ee
where $\zeta$ is not integrated over in (\ref{eq-mahler-m-ererg}).
Hence, the correspondence between the Ising model and spanning trees
generalizes to the triangular lattice (and very likely also 
to other planar lattices).  }

For nonplanar lattices, however, the issue is not at
all clear. Consider for example the simple cubic lattice.  The random
walk structure function and the spanning tree generating function are
known~\cite{rosengren,arXiv:1207.2815}.
Specifically, the structure function for the simple cubic lattice
is given by
\be
\Lambda(k_1,k_2,k_3)
={1\over 3} (\cos k_1+\cos k_2+\cos k_3)
~.
\ee
Hence, the spanning tree generating function for the cubic lattice
can be expressed in terms of the corresponding Mahler measure. 
The relevant Mahler measure is the one given by
choosing $P$ to be
the
symmetric 3-variable
Laurent polynomial
\be
s(\zeta)= \zeta + z_1 + z_1^{-1} +  z_2 + z_2^{-1}+  z_3 + z_3^{-1}
~.
\ee
It is possible to show that this Mahler measure cannot be
reconciled with the known series expansion of the partition function
of the cubic Ising model, in a manner analogous to the two dimensional
case.
Specifically, it is easy to check that if $\zeta$ is chosen to be an
arbitrary rational function in the high temperature variable $t=\tanh
\beta J$ then one cannot recover the known high temperature series
expansion for the partition function, unlike the case for the two
dimensional Ising model.
Hence, there can be no simple relation between the partition function
of the Ising model and the spanning tree generating function on the
cubic lattice.
(Indeed, this should be obvious, since otherwise the cubic Ising model
would have been solved long ago.)

Nevertheless, there are a number of related unanswered questions, for
instance, might it be possible to rule out all Mahler measures for
the Ising model on 
non-planar lattices? Currently, it is widely believed that the partition
functions for non-planar lattice models are not D-finite, whereas
Mahler measures by definition are D-finite.  Without proofs, of
course, such statements remain conjectural.  These and related
questions
questions remain open and the posibilities are intriguing. We believe
they merit further study and hope to investigate them in the future.

\section*{Acknowledgments}

We thank CNPq for funding. We thank MGE~da~Luz, RTG~de~Oliveira and
EP~Raposo for very helpful comments.

\appendix*

\section {Proof of Theorem 1 via   $_p F_q$ functions}

On the one hand,   it is known 
from Eq. (23) in ref.~\cite{arXiv:1207.2815}  that
  
\bel{eq-def-T-pfq}
T(z)=
\log \frac 4  z
-
{z^2\over 8}
{_4}F_3
\left[ \ba{c}{1,1,{3\over 2},{3\over 2}} \\ {2,2,2}\ea ;
z^2
\right] ~.
\ee
On the other hand, it is known from Eq. (13) in ref.~\cite
{viswan2015-jstat} that 

\begin{align}
\ln Z(K)
&= \ln(2 \cosh 2K) 
- \kappa^2 ~{_4}F_3
\left[ \ba{c}{1,1,{3\over 2},{3\over 2}} \\ {2,2,2}\ea 
; 16 \kappa^2 \right]  \label{eq-Z-pfqdef}
\end{align}
with  $\kappa$ defined as
\bel{eq-k-def}
2\kappa = \tanh(2K) \sech(2K)
 ~.
\ee

Putting $z=4 \kappa$ in (\ref{eq-def-T-pfq}) gives us 

\be
-\kappa^2 {_4}F_3
\left[ \ba{c}{1,1,{3\over 2},{3\over 2}} \\ {2,2,2}\ea ;
16\kappa^2
  \right]
=
\frac 1 2 
\big(
T(4\kappa)+ 
\log \kappa
\big)
~.
\ee
Substituting this last expression into (\ref{eq-Z-pfqdef}) to
eliminate the $_4 F_3$ function, and then exponentiating, we obtain 

\begin{align}
Z(K)
&=
2 \cosh 2K \kappa^{1/2}
\exp\left[\tfrac 1 2 
T(4\kappa)
\right]
~.
\end{align}
The claim follows from noting that
$4\kappa=k$.
$\square$

\end{document}